\documentclass[9pt,twocolumn,twoside]{optica}
\setboolean{shortarticle}{true}
\setboolean{minireview}{false}

\title{Photonic spin Hall effect in hyperbolic metamaterials at visible}

\author[1,*]{Osamu Takayama}
\author[1]{Johneph Sukham}
\author[1]{Radu Malureanu}
\author[1]{Andrei V. Lavrinenko}
\author[2,3]{Graciana Puentes}

\affil[1]{DTU Fotonik -- Department of Photonics Engineering, Technical University of Denmark, {\O}rsteds Plads 343, DK-2800 Kgs. Lyngby, Denmark}
\affil[2]{Universidad de Buenos Aires. Facultad de Ciencias Exactas y
Naturales. Departamento de F\'{\i}sica. Buenos Aires, Argentina}
\affil[3]{CONICET-Universidad de Buenos Aires. Instituto de F\'{\i}sica de Buenos Aires (IFIBA). Buenos Aires, Argentina}

\affil[*]{Corresponding author: otak@fotonik.dtu.dk, gpuentes@df.uba.ar}

\ociscodes{(160.3918) Metamaterials; (240.5440) Polarization-selective devices; (260.0260) Physical optics.}

\begin{abstract}
Photonic spin Hall effect in transmission is a transverse beam shift of the out-coming beam depending on polarization of the in-coming beam. The effect can be significantly enhanced by materials with high anisotropy. We report the first experimental demonstration of the photonic spin Hall effect in a multilayer hyperbolic metamaterial at visible wavelengths (wavelengths of 520 nm and 633 nm). The metamaterial is composed of alternating layers of gold and alumina with deeply-subwavelength thicknesses, exhibiting extremely large anisotropy. The angle resolved polarimetric measurements showed the shift of 165 $\mu m$ for the metamaterial of 176 nm in thickness. Additionally the transverse beam shift is extremely sensitive to the variations of the incident angle changing theoretically by 270 $\mu m$ with one milli-radian  ($0.057^\circ$). These features can lead to minituarized spin Hall switches and filters with high angular resolution.
\end{abstract}

\setboolean{displaycopyright}{true}

\begin{document}

\maketitle

The photonic spin Hall effect \cite{1} or spin Hall effect of light \cite{2} is the photonic analogue of the spin Hall effect occurring with charge carriers in solid-state systems. Typically this  phenomenon takes place when a light beam refracts at an air-glass interface, or when it is projected onto an oblique plane, the latter effect being known as the geometric spin Hall effect of light. In general the photonic spin Hall effect leads to a polarization-dependent transverse y-shift of a light peak intensity \cite{3,4,5,7}. An example of the latter effect is the transverse Imbert-Federov beam shift \cite{3}, which happens for paraxial beams reflected or refracted at a sharp inhomogeneity of an isotropic optical interface 
\cite{4, 5, 7, 8}. Potential applications of the photonic spin Hall effect in spin-dependent beam splitters, optical diodes \cite{9} and surface sensors \cite{10,11} are considered in various fields in photonics such as nanophotonics, plasmonics, metamaterials, topological optics and quantum optics \cite{12,10}.

Photonic spin Hall effect has been studied in reflection and transmission with various materials and geometrical settings.  In the reflection configuration, the effect has been studied experimentally on interfaces of  uniaxial dielectrics (LiNbO$_3$) \cite{12}, BK-7 glasses \cite{13}, glass prisms \cite{14}, metal (Ag) films \cite{15}, magnetic films \cite{16}, dielectric multilayers \cite{17} and topological insulators \cite{18}. In the case of the transmission configuration, the photonic spin Hall effect has been experimentally demonstrated in quarts crystals \cite{19}, anisotropic polymers \cite{20}, metal (Au) films \cite{21}, liquid crystals \cite{22} and dielectric spheres \cite{23}.

\begin{figure}[htbp]
\centering
\fbox{\includegraphics[width=0.9\linewidth]{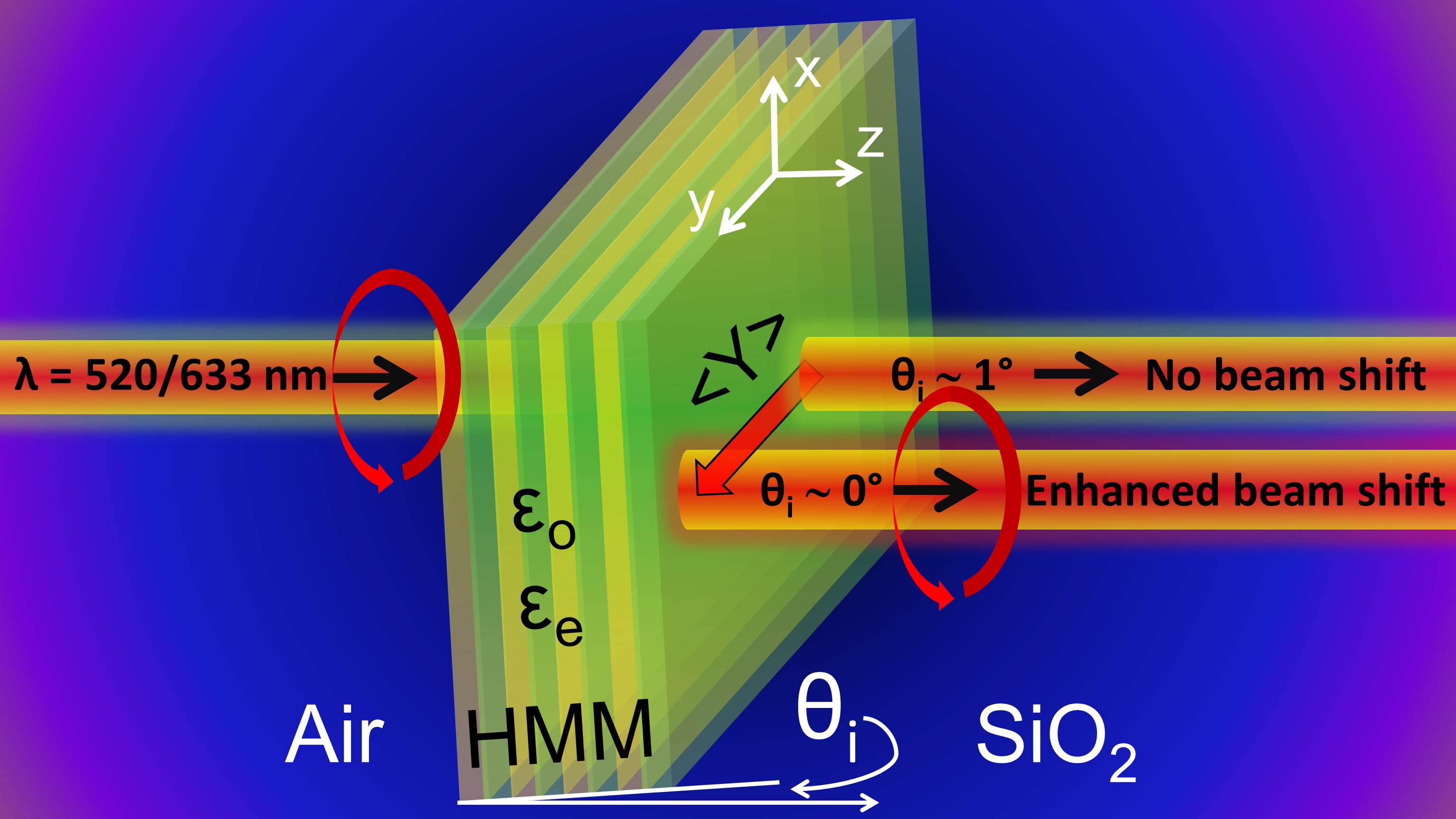}}
\caption{Schematic illustration of the spin Hall effect. The transverse beam shift along the y-axis denoted as $<Y>$ is switched by incident angle $\theta_i$. The unit cell of the HMM structure consists of Al$_2$O$_3$(10 nm)-APTMS(1 nm)-Au(10 nm)-APTMS(1 nm). The HMM contains eight periods  with the total thickness of 176 nm.}
\label{fig:fig1}
\end{figure}

Recently, metasurfaces and metamaterials, artificially engineered subwavelength surface and volume structures, have stimulated significant interest thanks to their flexible design parameters \cite{24}. A subclass of metamaterials - hyperbolic metamaterials (HMM) exhibit extreme anisotropy with hyperbolic dispersion profiles in the wavevector space. Conventionally, HMMs take the form of a multilayer of alternating metal and dielectric thin films \cite{25,26,27}, trenches \cite{29,30,31} or metallic nanowires \cite{32,33}.
In the typical cases HMMs possess uniaxial anistropy, and their optical properties are characterized by effective ordinary and extraordinary permittivities, $\varepsilon_{o}$ and $\varepsilon_{e}$, respectively, as depicted in Fig \ref{fig:fig1}. Depending on the sign of $\varepsilon_{o}$ and $\varepsilon_{e}$, a metamaterial can be categorized as so-called type I ($\varepsilon_{o}>0$ and $\varepsilon_{e}<0$) or type II HMM ($\varepsilon_{o}<0$ and $\varepsilon_{e}>0$) \cite{34,35}.  Such unique optical properties of HMMs have led to various applications, sub-diffraction imaging \cite{36} and sensing \cite{37,38,39}, to name a few.

The spin Hall effect in HMMs has been theoretically studied recently \cite{40,41}. Experimental observation of the spin Hall effect in HMMs has been reported so far only for microwave frequencies \cite{42}. The hyperbolic dispersion obtained with electronic components, such as capacitors and inductors, helps to steer directional surface waves by input polarization states.

Here, we experimentally demonstrate extreme angle sensitivity and enhancement of the spin-Hall effect of light in HMMs in the visible regime. The effect is shown in the transmission configuration with a few hundreds nanometers thick HMM composed of alternating layers of metal and dielectric as illustrated in Fig. \ref{fig:fig1}. The transverse beam shift in our HMM setting is very sensitive to the angle of incidence, drastically changing from almost no beam shift to a few hundred microns by the change of $\approx$ 0.003 rad ($\approx$ 0.17 $^\circ$) in the incident angle. The large photonic spin Hall enhancement in a such thin structure and extreme angular sensitivity can be exploited in compact spin Hall devices which enable manipulation of photons by polarization. 

The HMM sample with the total thickness of 176 nm  contains eight gold-alumina periods deposited on a 500 $\mu$m thick glass substrate. More detailed, one period  of the HMM structure  consists of four layers: Al$_2$O$_3$(10 nm)--APTMS(1 nm)--Au(10 nm)--APTMS(1 nm).  APTMS, which stands for Amino Propyl Tri Methoxy Silane, is an almost loss-free adhesion layer advantageous for highly localized propagating plasmon modes \cite{27}. The Au layer was sputtered, and Al$_2$O$_3$ layer was deposited by atomic layer deposition. 
The treatment of HMMs as homogenized uniaxial media with effective permittivities are based on 
the effective media approximation \cite{43}, which is assumed to be valid under the condition that the thicknesses of individual layers are deeply subwavelength.   The unit cell is $\Lambda$ = 22 nm, and normalized by the wavelength interval of $\lambda$ = 500 -- 700 nm, provides the ratio of $\Lambda/\lambda$ = 1/22.7 -- 1/31.8. Thus application of  the effective media approximation is our case is justified. 

\begin{figure}[htbp]
\centering
\fbox{\includegraphics[width=0.7\linewidth]{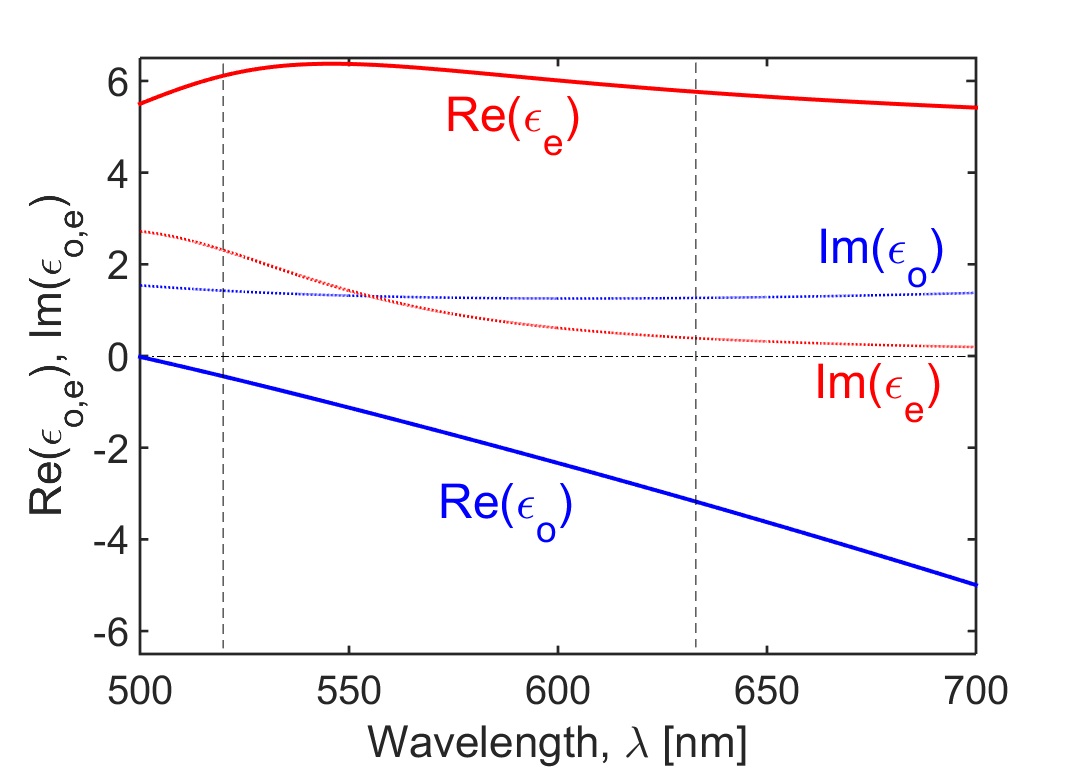}}
\caption{Effective permittivities of multilayer HMM structure composed of a unit cell of Al$_2$O$_3$(10 nm)-APTMS(1 nm)-Au(10 nm)-APTMS(1 nm) calculated by the effective media approximation.  In the visible wavelength interval, the structure behaves as type II ($\varepsilon_o <$ 0 and $\varepsilon_e >$ 0) HMM.  The wavelengths $\lambda$ = 520 nm and 633 nm at which we conducted experiments are noted by the vertical dashed lines.}
\label{fig:fig2}
\end{figure}

The ordinary and extraordinary permitivities of the HMM, denoted as $\varepsilon_{o}$ and $\varepsilon_{e}$ respectively, are calculated as \cite{30} 

\begin{equation}
\varepsilon_{o}=f_{Au}\cdot\varepsilon_{Au}+f_{Al_2O_3}\cdot\varepsilon_{Al_2O_3}+f_{APTMS}\cdot\varepsilon_{d_{APTMS}}
\end{equation}

\begin{equation}
\varepsilon_{e}=(\frac{f_{Au}}{\varepsilon_{Au}}+\frac{f_{Al_2O_3}}{\varepsilon_{Al_2O_3}}+\frac{f_{APTMS}}{\varepsilon_{APTMS}})^{-1},
\end{equation}

\noindent where $\varepsilon_{m}$ and $\varepsilon_{d}$ are the permittivities of metal and dielectric, and $f_{m}$ and $f_{d}$ are the volume fractions of metal and dielectric, respectively. The permittivity of the Au film, $\varepsilon_{m}$, is characterized by the Drude--Lorentz model with the thickness-dependent correction \cite{31}.  
The refractive index of APTMS is 1.46 \cite{32}. 
Fig. \ref{fig:fig2} shows dispersion of the HMM effective permittivities in the visible range. Our HMM structure has a zero crossing wavelength for $\varepsilon_o$ around $\lambda$ = 500 nm and further to the red wavelengths becomes type II HMMs ($\varepsilon_o <$ 0 and $\varepsilon_e >$ 0).

In order to understand the behavior of the spin Hall beam shift in HMMs, we conducted simulation based on a theory developed by T. Tang \textit{et al. }\cite{43,44} with realistic parameters of fabricated HMMs presented in Fig. \ref{fig:fig2}. The whole HMM-SiO$_2$ substrate structure has air as the ambient medium as shown in Fig. \ref{fig:fig1}. We assume the incident light is impinging on the HMM structure in the y-z plane with incident angle $\theta_{i}$. The relative permittivities of the media in regions 1-5 are denoted by $\epsilon_{i}$ $(i = 1, 2, 3, 4, 5)$ respectively, where  $\epsilon_1=\epsilon_2=\epsilon_5$ = 1 (air), $\epsilon_4$ corresponds to the SiO$_2$ substrate and $\epsilon_3$ corresponds to the HMM. The HMM is assumed to be non-magnetic and uniaxial anisotropic with a relative permittivity tensor ($\epsilon_3$):

\begin{equation}
\epsilon_3 = 
\left(\begin{array}{ccc} \epsilon_{o} & 0& 0\\ 0 & \epsilon_{o} & 0  \\
0 & 0 & \epsilon_{e}
\end{array}\right)
\end{equation}

Considering the input Gaussian beam of waist $\omega_0$:
\begin{equation}
E_{H,V}(x,y)=\frac{\omega_0}{\sqrt {2 \pi}}\exp{-\frac{\omega_0^2(k_{x}^2+k_{y}^2)}{4}}
\end{equation}
we can define the transverse beam shifts after transmission through the structure in the form:
\begin{equation}
\eta_{H,V}^{\pm}=\frac{x|E_{H,V}|^{\pm}}{|E_{H,V}|^{\pm}},
\end{equation}
\noindent where $\eta^{\pm}$ indicate transverse shifts for the right hand circular (RHC) and left hand circular (LHC) polarizations. The transverse shifts include z-dependent and z-independent terms which represent spatial and angular transverse shifts, respectively \cite{33}. Here, we focus on the spatial transverse shift of transmitted light through the HMM waveguide, which takes the form \cite{34}:
\begin{equation}
<Y>=\pm \frac{k_1 \omega_0^2(t_{s}^2\frac{\cos (\theta_{t})}{\sin (\theta_{i})}-t_{s}t_{p}\cot (\theta_{i}))}
{k_{1}^{2}\omega_0^2t_{s}^2+\cos^2(\theta_{i})(t_{s} \frac{\cos (\theta_{t})}{\cos(\theta_{i})}-t_{p})^2+ (\frac{d t_{s}}{d \theta_{i}} )^2  },
\end{equation}
\noindent where $k_{1}=n_1 k=n_1\frac{2 \pi}{\lambda}$ with $n_1=1$ (air), $t_{s,p}$ are the transmission amplitudes for the $s,p$ modes, respectively \cite{43}, $\theta_{t}$ is the transmission angle.  We consider $(\frac{d t_{s}}{d \theta_{i}}) \approx 0$ for a large beam waist and $\theta_t$ = 0, assuming transmission along the laser beam axis.

\begin{figure}[htbp]
\centering
\fbox{\includegraphics[width=0.95\linewidth]{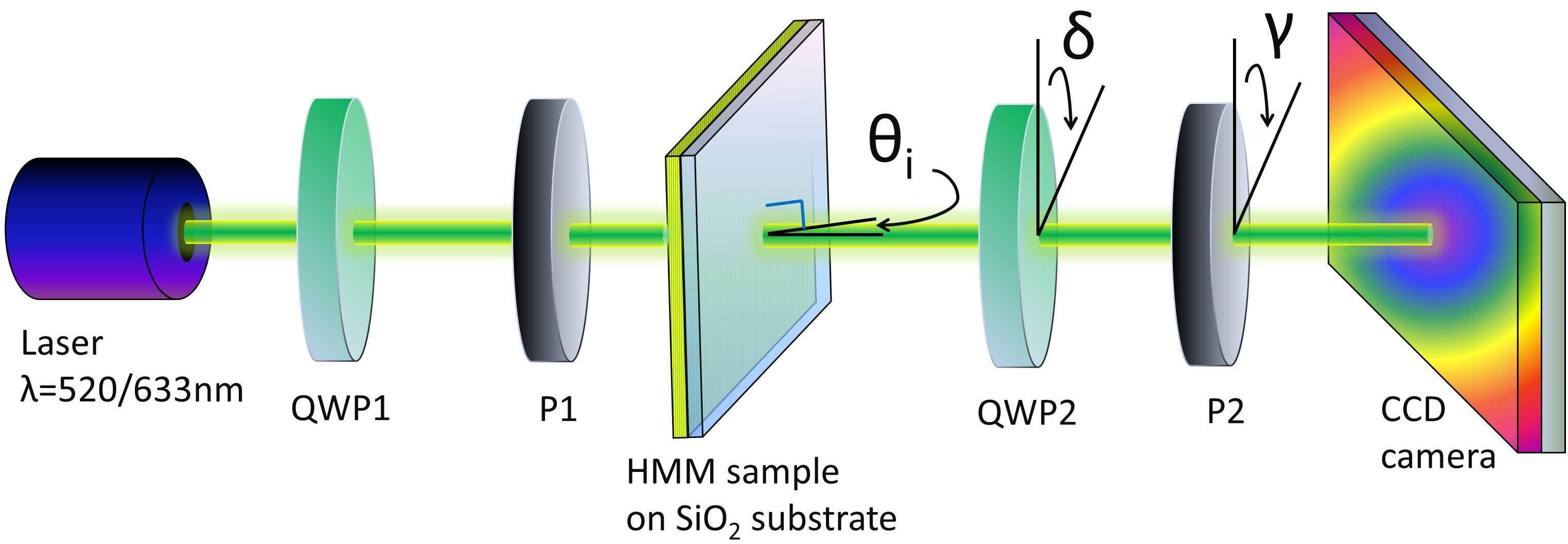}}
\caption{Schematic illustration of polarimetric measurement setup.  P1 and P2 represent double Glan-Laser polarizers (Thorlabs GL10), QWP1 and QWP2 are quarter wave plates. Laser is either a He-Ne laser (Melles Griot 05-LHR-111) or diode laser (Thorlabs LP 520-MF-100) with wavelength of 633 nm and 520 nm, respectively. The CCD camera is Thorlabs WFS150-5C.}
\label{fig:Setup}
\end{figure}

To demonstrate the angular sensitivity of the photonic spin Hall shift in the HMM, we performed a series of characterizations using the polarimetric setup shown in Fig. \ref{fig:Setup}. 
The transverse beam shift can be measured by polarimetric and quantum weak measurement \cite{21,22,45}. As a source of an incident Gaussian beam, we employed a He-Ne laser with wavelength $\lambda = 633$ nm, and a diode laser with wavelength $\lambda =520$ nm. The laser radiation was collimated using a microscope objective lens. We measured the anisotropic phase difference $\Phi_0$ versus angle $\theta_i$ with the Stokes polarimetry method \cite{21}. The input polarization state (RHC) is prepared using a Quarter Wave-Plate (QWP) followed by a Glan-Thompson polarized (P1). The phase difference can be obtained via the Stokes parameters using the expression:

\begin{equation}
\Phi_0=\arctan (S_3/S_2),
\label{phase}
\end{equation}

\noindent where $S_{3}=I(90 ^\circ,45 ^\circ)-I(90 ^\circ,135 ^\circ)$  is the normalized Stokes parameter for the circular polarization, and $S_{2}=I(0 ^\circ,45 ^\circ)-I(0 ^\circ,135 ^\circ)$  is the normalized Stokes parameter in the diagonal basis, where normalization factor $S_0$ is given by the total intensity of the beam. $\delta$ and $\alpha$ of $I(\delta,\alpha)$ correspond to the retardation angle of quarter wave plate QWP2, and the rotation angle of polarizer P2, respectively. The measured phase using Eq.\ref{phase} is wrapped in the range (-$\pi,\pi$). In order to determine the unwrapped phase difference we use the unwrapping algorithm \cite{21},
with a tolerance set to 0.01 rad. 

The transverse beam shift, $<Y>$, is found from the measured phase, $\Phi_0$,  
\begin{equation}
<Y> = \frac{1}{k}cot(\pi/2-\theta_i)[-\sigma(1-cos\Phi_0)+\chi sin\Phi_0],
\label{eq_kY}
\end{equation}
\noindent where $\sigma$ and $\chi$ are the Stokes parameters in the circular and diagonal basis, respectively. They are given by $\sigma$ = 2Im($\alpha^*\beta$), and $\chi$ = 2Re($\alpha^*\beta$) from the Jones vector $|\psi\rangle=(^\alpha_\beta)$ of the incident beam, respectively. The incident beam has the RHC polarization, that is, $\alpha= \frac{1}{\sqrt{2}},\beta=\frac{i}{\sqrt{2}}$.

\begin{figure}[htbp]
\centering
\fbox{\includegraphics[width=0.82\linewidth]{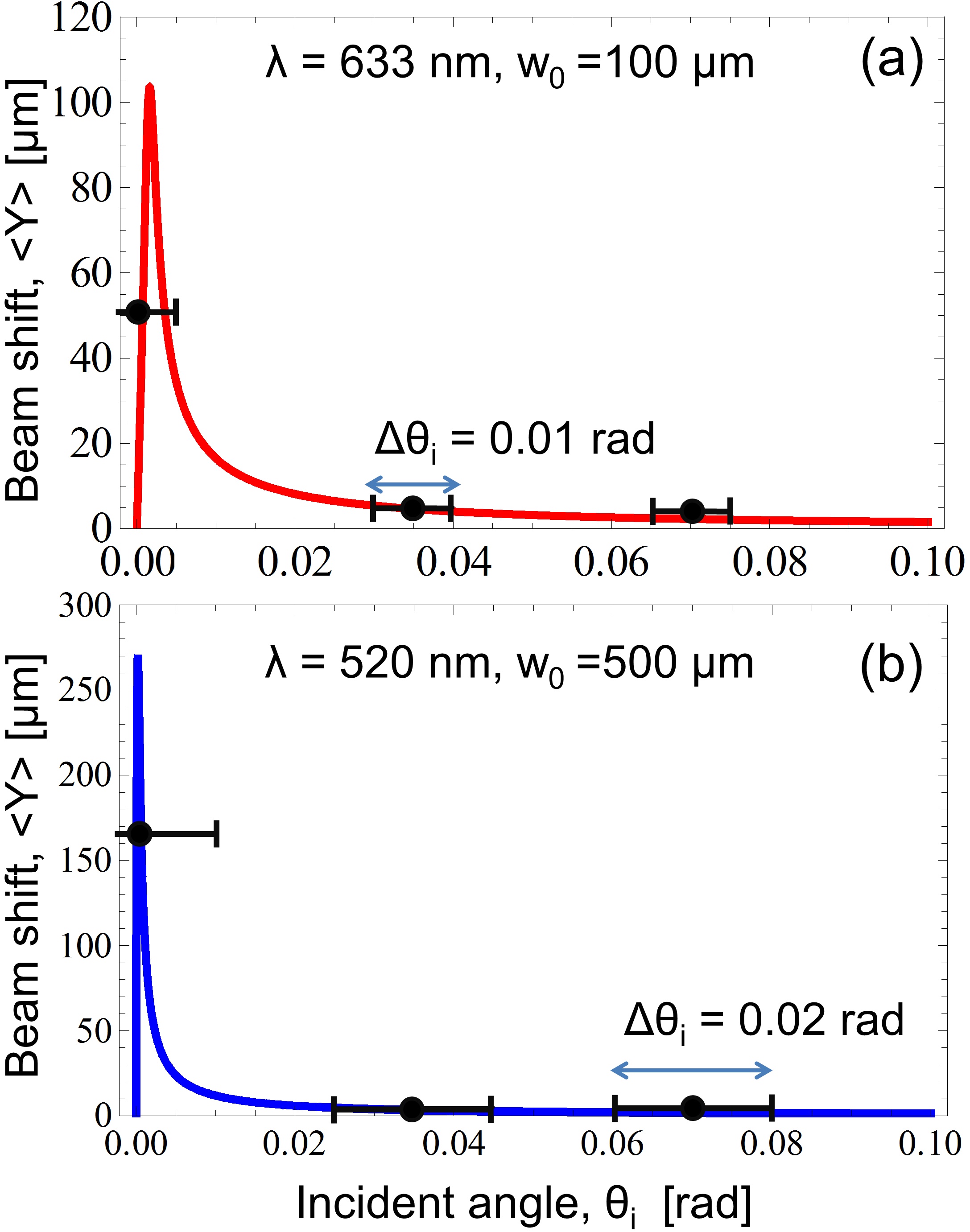}}
\caption{Measured and simulated (solid lines) spin Hall transverse shift, $<Y>$, for (a) $\lambda$ = 633 nm and (b) $\lambda$ = 520 nm, respectively. Simulation results in (a) and (b) are fitted in terms of beam width $w_o$. Note that there is an estimated beam divergence of $\Delta\theta_i$ = 0.01 rad for $\lambda$ = 633 nm  and $\Delta\theta_i$ = 0.02 rad for $\lambda$ = 520 nm indicated as lateral error bars, respectively.}
\label{fig:fig5}
\end{figure}

Fig. \ref{fig:fig5} shows the experimental points of the beam shifts for wavelengths $\lambda$ = 520 nm and $\lambda$ = 633 nm versus the simulated results, where losses of the HMM were taken into account. The effective permittivitties are $\varepsilon_o$ = -2.9460 + 1.2163i and $\varepsilon_e$ = 5.5063 + 0.3399i for $\lambda$ = 633 nm, and $\varepsilon_o$ =  - 0.3257 + 1.3664i and $\varepsilon_e$ = 5.8847 +  1.9963i for $\lambda$ = 520 nm, respectively, see Fig. \ref{fig:fig2}.  
Here we can observe the peculiarities of the photonic spin Hall effect in HMM structures in the transmission configuration.  The tranverse beam shift in HMMs is extremely sensitive to the incident angle: its variation from $\theta_i$ = 0 rad to merely $\theta_i$ = 0.003 rad (0.17 $^\circ$) for $\lambda$ = 633 nm [Fig. \ref{fig:fig5}(a)] induces the giant beam shift of a few hundreds microns, exhibiting sensitivity in the milli-radian range. The angular variation from $\theta_i$ = 0.003 rad to above also significantly changes the beam shift from $<Y>$ = 105 $\mu$m to merely $<Y>$ = 10 $\mu$m, what is almost one order of magnitude difference [Fig. \ref{fig:fig5}(a)]. The sharp peak of the beam shift is attributed to large anisotropy of HMMs, $\varepsilon_e - \varepsilon_o$, as we also saw such behavior in dielectric media \cite{22}. In the case of $\lambda$ = 520 nm with a larger beam diameter [Fig. \ref{fig:fig5}(b)], the beams shift exhibits even sharper resonance and, consequently, enhanced angular sensitivity: the peak shift of $<Y>$ = 270 $\mu$m is achieved by the incident angle change of 0.001 rad (0.057 $^\circ$) only ($\approx$ 4700 $\mu$m/$^\circ$).  When the incident angle becomes larger, for instance, $\theta_i$ = 0.01 rad (0.57 $^\circ$), the beam shift drastically drops to $<Y>$ = 10 $\mu$m and less. 

The experimental results are in the full quantitative agreement with the simulated values taking into account beams divergence and setups errors. The beam divergence of the He-Ne laser is estimated to be about $\Delta\theta_i$ = 0.01 rad (0.57 $^\circ$) and of the green diode laser is about $\Delta\theta_i$ = 0.02 rad (1.14 $^\circ$). Additionally, the angular resolution is limited by the rotating mount with error about 0.0017 rad (0.1 $^\circ$).  For $\lambda$ = 633 nm [Fig. \ref{fig:fig5}(a)], the experimental transverse beam shift under the normal incidence reaches $<Y>$ = 50 $\mu$m.  However, when the incident angle is tilted by merely $\theta_i$ = 0.035 rad ($\approx$ 2 $^\circ$), the transverse beam shift dramatically drops down to $<Y>$ = 5 $\mu$m. Such results confirm that the transverse beam shift can be tuned within one order of magnitude range by small angular variations demonstrating extreme angular sensitivity of the spin Hall effect in HMMs. 
The waist used for the simulations ($w_0$ = 100 $\mu$m) is measured experimentally.  

In the case of $\lambda$ = 520 nm, we observe $<Y>$ = 165 $\mu$m for $\theta_i$ = 0 rad as shown in Fig. \ref{fig:fig5}(b). At $\theta_i$ = 0.035 rad ($\approx$ 2 $^\circ$), the beam shift decreases to $<Y>$ = 5 $\mu$m. This extremely high angular sensitivity makes a striking contrast to previously observed spin Hall effect in dielectric anisotropic media, such as a quartz crystal ($\approx$ 150 $\mu$m / 5 $^\circ$ = 30 $\mu m/^\circ$) \cite{21} and polymer film ($\approx$ 250 $\mu$m / 20 $^\circ$ = 12.5 $\mu m/^\circ$) \cite{22}, where the beam shift of the same magnitude occurs within several degrees of sample tilting, that is two orders of magnitude larger.  Moreover, the thickness of the HMM is 176 nm only as opposed to those of the dielectric materials: $50 \mu m$ for the polymer film and $1 mm$ for the quartz plate.  

In conclusion, we experimentally demonstrated for the first time the photonic spin Hall effect in a hyperbolic metamaterial at visible wavelengths. The tranverse beam shift in the transmission configuration is very sensitive to the incident angle: we observed that a few milliradian difference changes the beam shift by two orders of magnitude, for example, from a few hundreds of microns down to several microns. This extreme angular tunability is realized in a two hundreds of nanometers thick HMM. Such sensitivity can lead to thin and compact spin Hall devices that manipulate light at nanoscale by means of spin, incident angle, and wavelength, such as switches, filters, sensors. 

\bigskip
\noindent\textbf{Funding Information.}
Agencia Nacional de Promocion Cientifica y Tecnologica, PICT2015-0710 Startup,  UBACyT PDE 2016, UBACyT PDE 2017, Argentina; Villum Fonden DarkSILD project (11116); Direkt{\o}r Ib Henriksens Fond, Denmark.

\bigskip
\noindent\textbf{Acknowledgments.}
The authors are grateful to Konstantin Bliokh, Mark Dennis, and Joerg Goette for fruitful discussions. 





\end{document}